\documentclass[10pt]{iopart}
\makeatletter
\@namedef{ver@amsmath.sty}{}
\makeatother
\usepackage{amstext}
\usepackage{graphics}
\usepackage{graphicx}
\usepackage{cite}
\usepackage{epsfig}
\usepackage{epstopdf}
\usepackage{subfigure}
\usepackage{braket}
\DeclareMathAlphabet{\mathpzc}{OT1}{pzc}{m}{it}
\usepackage[dvips]{color}
\usepackage{gensymb}
\usepackage{dcolumn}
\usepackage{bm}
\usepackage{amssymb}
\usepackage{iopams}
\usepackage{amsmath}
\usepackage{indentfirst}
\usepackage{booktabs}
\usepackage{multirow}
\usepackage{colortbl}

\begin{document}

\title{Antiferromagnetism-induced second-order nonlinear optical responses of centrosymmetric bilayer CrI$_3$}
\author{Vijay Kumar Gudelli$^1$ and Guang-Yu Guo$^{2,1}$}
\address{$^1$ Physics Division, National Center for Theoretical Sciences, Hsinchu 30013, Taiwan}
\address{$^2$ Department of Physics and Center for Theoretical Physics, National Taiwan University, Taipei 10617, Taiwan}

\ead{gyguo@phys.ntu.edu.tw}

\vspace{10pt}
\date{\today}

\begin{abstract}
Antiferromagnetism (AF) in AB$'$-stacked centrosymmetric bilayer (BL) CrI$_3$ 
breaks both spatial inversion ($P$) and time-reversal ($T$) symmetries
but maintains the combined $PT$ symmetry, thus inducing novel second-order
nonlinear optical (NLO) responses such as second-harmonic generation (SHG),
linear electric-optic effect (LEO) and bulk photovoltaic effect (BPVE).
In this work, we calculate AF-induced NLO responses of
the BL CrI$_3$ based on the density
functional theory with the generalized gradient approximation (GGA) plus 
onsite Coulomb correlation (U), i.e., the GGA+U method.
Interestingly, we find that the magnetic SHG, LEO and photocurrent in the AF
BL CrI$_3$ are huge, being comparable or even larger than
that of the well-known nonmagnetic noncentrosymmetric semiconductors.
For example, the calculated SHG coefficients are in the same order of magnitude
as that of MoS$_2$ monolayer (ML), the most promising 2D material for NLO devices.
The calculated LEO coefficients are almost three times larger than
that of MoS$_2$ ML. The calculated NLO photocurrent in the CrI$_3$ BL
is among the largest values predicted so far for the BPVE materials.
On the other hand, unlike nonmagnetic semiconductors, the NLO responses in the AF BL CrI$_3$ are
nonreciprocal and also switchable by rotating magnetization direction.
Therefore, our interesting findings indicate that the AF BL CrI$_3$ will not
only provide a valuable platform for exploring new physics 
of low-dimensional magnetism but also have promising applications in magnetic
NLO and LEO devices such as frequency conversion, electro-optical switches,
and light signal modulators as well as high energy conversion efficiency
photovoltaic solar cells.
\end{abstract}

%
\ioptwocol

\section{Introduction}

Recently, two groups discovered long-range magnetic orders in atomically thin films of semiconductors CrGeTe$_3$ \cite{gong2017} 
and CrI$_3$ \cite{huang2017}. This exciting discovery has opened up new research directions for two-dimensional (2D) materials. 
In particular, magnetism at 2D limit is highly desirable for both the fundamental physics and also for the technological 
applications ranging from magnetic memories to sensing, to spintronics to novel functionalities based on 2D materials. 
Consequently, magnetic 2D materials are currently subject to intensive investigations.
Among the magnetic 2D materials, bilayer (BL) CrI$_3$ seems to be unique and attracts particularly
strong attention \cite{seyler-nat2018,jiang-nm2018,jiang-nnt2018,huang-nnt2018,kim-nl2018,klein-sc2018,song-sc2018}.
Unlike other atomically thin films of CrGeTe$_3$ and CrI$_3$ which exhibit a ferromagnetic 
long-range order,~\cite{gong2017,huang2017}, BL CrI$_3$ shows an interlayer antiferromagnetic (AF) order~\cite{huang2017}.
This AF order in BL CrI$_3$ leads to a number of emerging phenomena, such as giant tunnelling magnetoresistance 
in spin-filter tunnel junctions~\cite{song-sc2018,klein-sc2018} and the electrical control of 2D magnetism~\cite{jiang-nm2018,huang-nnt2018,jiang-nnt2018}.
First-principles density functional theory studies\cite{sivadas-nl2018,jiang-prb2019,soriano-ssc2019,jang-prm2019}
 indicated that the interlayer magnetic coupling in BL CrI$_3$ 
is highly stacking dependent \cite{sivadas-nl2018,jiang-prb2019,soriano-ssc2019,jang-prm2019} and that
the unusual AF coupling originates from the monoclinic AB$'$ stacking [see Fig. 1(c)] in BL CrI$_3$ instead of the usual trigonal AB
stacking [see Fig. 1(b)] in other CrI$_3$ multilayers. 

Very recently, giant nonreciprocal second-harmonic generation (SHG) was observed in the AF BL CrI$_3$\cite{sun-nature2019}.
SHG converts two incident photons of the same frequency ($\omega$) into a new photon with a doubled frequency ($2\omega$), 
and is a second-order nonlinear optical phenomenon \cite{shen_2003,boyd_2003}.
SHG plays an important role in the modern optics and electro-optical devices such as lasers,
electro-optical modulators and switches, frequency conversions \cite{shen_2003,boyd_2003}. 
Second-order NLO phenomena occur in nonmagnetic materials with the broken spatial inversion ($P$) symmetry.~\cite{shen_2003,boyd_2003}
The AB$'$ stacked BL CrI$_3$ has the centrosymmetric $C_{2h}$ symmetry group [see Fig. 1(c)], and nonmagnetic
AB$'$ stacked BL CrI$_3$ would not exhibit second-order NLO responses. 
Intriguingly, the antiferromagnetism in the AB$'$ stacked BL CrI$_3$ breaks both time-reversal ($T$)
and spatial inversation symmetries \cite{sun-nature2019} [see Fig. 1(d)] but maintains 
the combined $PT$ symmetry~\cite{Zhang2019a}.
This results in a number of emergent magnetic NLO responses such as magnetic SHG reported in Ref. \cite{sun-nature2019}
and opens up possibilities for the application of magnetic 2D materials in magnetic controllable 
NLO and nonreciprocal optical devices.

In this work, we calculate three AF-induced NLO responses of
the BL CrI$_3$, namely, SHG, linear electric-optic effect (LEO)
and nonlinear injection photocurrent. Our calculations are based on the density
functional theory with the generalized gradient approximation (GGA) plus
onsite Coulomb correlation (U), i.e., the GGA+U method.
Interestingly, we find that the present calculated magnetic SHG, LEO and photocurrent in the AF
BL CrI$_3$ are huge in comparison with the well-known nonmagnetic noncentrosymmetric semiconductors with the same properties.
In particular, the calculated SHG coefficients of the AF BL CrI$_3$ are in the same order of magnitude
as that of MoS$_2$ monolayer (ML), which is the best candidate of the NLO material among the 2D materials. 
Further, the calculated LEO coefficients derived from SHG coefficients, are almost three times higher than 
the reported LEO coefficients of MoS$_2$ ML. 
We observe that the NLO photocurrent in BL CrI$_3$ is robust and it is one of the highest values predicted so far for the BPVE materials.
Interestingly, we also find that the NLO responses in the AF BL CrI$_3$ are nonreciprocal and also reversible by rotating magnetization direction, 
which is very different from the nonmagnetic NLO semiconductors.
This work thus demonstrates that because of its novel AF-induced NLO properties, AF BL CrI$_3$ 
will find promising applications in magnetic based NLO devices scuh as electro-optical switches,
frequency conversion and light signal modulators as well as high energy conversion efficiency
photovoltaic solar cells.

\begin{figure}[htb]
\begin{center}
\includegraphics[width=7cm]{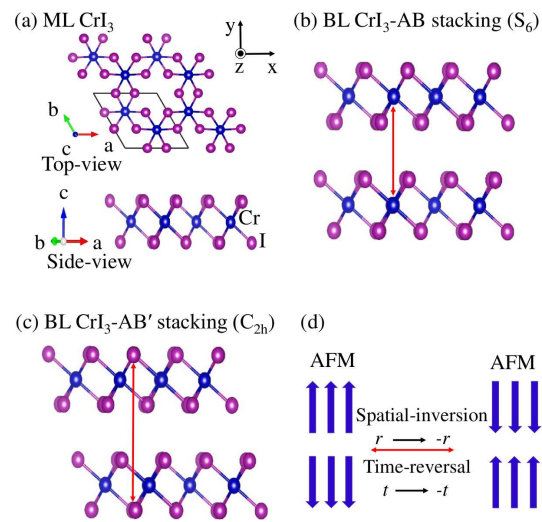}
\end{center}
\caption{(a) Top and side view of ML CrI$_3$.  The black solid line denotes the primitive cell. 
(b) The AB stacked BL CrI$_3$ with rhombohedral symmetry of $S_6$ point group. 
(c) The AB' stacked BL CrI$_3$ with monoclinic symmetry of $C_{2h}$ point group. 
This structure can be obtained by a lateral shift of bottom layer 
of the AB stacked BL CrI$_3$ in (b). (d) Schematic representation of the interlayer 
AF configuration in the AB$'$ stacked BL CrI$_3$ and the effects of the $P$ and $T$ symmetry operations. 
}
\end{figure}

\section{Theory and computational methods}
Bulk CrI$_3$ forms a layered structure with the monolayers (MLs) separated by the van der Waals gap. 
Each CrI$_3$ ML consists of edge-sharing CrI$_6$ octahedra
forming a planar network with Cr atoms in a honeycomb lattice [Figs 1(a)].
These MLs are then stacked in an ABC sequence, resulting in a rhombohedral crystal with $R\bar{3}$ symmetry.  
This structure can also be regarded as an ABC-stacked hexagonal crystal
with experimental lattice constants $a=6.867$ \AA$ $ and $c=19.807$ \AA~\cite{mag2015}.
For the CrI$_3$ BL structure, the two CrI$_3$ MLs can be stacked in two ways,
namely, the AB stacking with $S_6$ symmetry and the AB$'$ with $C_{2h}$ symmetry with the $y$-axis being the in-plane $C_2$ axis, 
as shown in Figs. 1(b) and 1(c), respectively. The monoclinic AB$'$ stacking is 
constructed using the AB stacked BL by a lateral shift of the bottom CrI$_3$ ML.
We have studied the magnetic and magneto-optical properties of the AB stacked CrI$_3$ BL 
before~\cite{vkg-njp2019}. Zhang {\it et al.} have calculated the NLO photocurrents 
in the AB stacked CrI$_3$ BL ~\cite{Zhang2019a}.
In this paper, we consider only the monoclinic AB$'$ stacked CrI$_3$ BL in the layered 
antiferromagnetic structure [(see Fig. 1(d)], since it is the experimentally observed 
crystalline structure for the AF BL CrI$_3$~\cite{sun-nature2019}.
Furthermore, we have calculated the total energies of the intralayer FM as well as
intralayer AF-Neel, AF-zigzag and AF-stripe magnetic configurations~\cite{vkg-njp2019},
and we found that all the intralayer AF structures have a total energy much higher than
that of the FM structure (at least 280 meV/cell higher), indicating that the
intralayer AF structures are unlikely to occur.
The slab-superlattice approach is used to simulate the BL structure with the
separation of the neighboring BLs being 15 \AA.

The electronic band structure, magnetic and optical properties of BL  CrI$_3$ are calculated based on first-principles 
density functional theory with the GGA~\cite{pbe_1996}. 
To better describe the Coulomb correlation among the Cr 3$d$ electrons, we adopt the GGA+$U$ scheme \cite{dudarev98}.
Following the recent studies \cite{Fang2018,vkg-njp2019}, we use $U = 1.0$ eV in the present calculations. 
The calculations are performed using the accurate projector-augmented wave (PAW) method~\cite{Bloechl1994,kresse1999ultrasoft},
as implemented in the Vienna {\it ab-initio} simulation package (VASP)~\cite{kresse1996efficient,kresse1996efficiency}.
The fully relativistic PAW potentials are adopted in order to include the SOC effect. 
A large plane-wave energy cutoff of 400 eV is used in all the calculations. 
For the Brillouin zone integration, a $k$-point mesh of 20 $\times$ 20 $\times$ 1 is used.
All the calculations are performed with an energy convergence within 10$^{-6}$ eV between the successive iterations.

The optical properties are calculated based on the linear response formalism with the 
the independent-particle approximation (IPA) in which the interaction between the electromagnetic field
and the electrons in the solid is treated as a perturbation. Aversa and Sipe have developed a
compact length-gauge formulation for calculating the nonlinear optical properties
such as the SHG, shift and injection currents of general solids including magnetic materials~\cite{Aversa1995}. 
Here we use the slightly re-arrangemed expressions from Ref. ~\cite{Aversa1995}. In particular,
the second-harmonic polarization density is given 
by $P_a^{(2)}(\omega)=\chi^{(2)}_{(abc)}(-2\omega;\omega,\omega)\mathcal{E}_b(\omega)\mathcal{E}_c(\omega)$
where $\mathcal{E}_b(\omega)$ denotes the cartesian $b$-component of the optical electric field 
and $\chi^{(2)}_{abc}(-2\omega,\omega,\omega)$ is the SHG susceptibility. 
The  $\chi^{(2)}_{(abc)}(-2\omega;\omega,\omega)$ can be written as 
\begin{equation}
\fl\chi^{(2)}_{abc}(-2\omega;\omega,\omega) = \chi^{(2)}_{abc,i}(-2\omega;\omega,\omega)
+\chi^{(2)}_{abc,e}(-2\omega;\omega,\omega),
\end{equation}
where 
\begin{eqnarray}
\chi^{(2)}_{abc,e} =&\frac{e^3}{\varepsilon_0 \hbar^2 V_c} \sum _{n,m,l}\sum _{\bf{k}} \frac{r_{nm}^a}{\omega_{mn}-2\omega}
(\frac{r_{ml}^br_{ln}^af_{nl}}{\omega_{ln}-\omega} \nonumber\\
&-\frac{r_{ml}^cr_{ln}^bf_{lm}}{\omega_{ml}-\omega})
\end{eqnarray}
is the contribution of the purely interband processes and 
\begin{eqnarray}
\chi^{(2)}_{abc,i} =&\frac{ie^3}{\varepsilon_0 \hbar^2 V_c} \sum _{n,m}\sum _{\bf{k}} 
(\frac{f_{nm}r_{nm}^a}{\omega_{mn}-2\omega}
(\frac{r_{mn;b}^c}{\omega_{mn}-\omega}-\frac{r_{mn}^c\Delta_{mn}^b}{\omega_{mn}-\omega}) \nonumber\\
&-\frac{f_{nm}r_{mn}^c}{2(\omega_{mn}-\omega)}(\frac{r_{nm;a}^b}{\omega_{mn}+\omega}-\frac{r_{nm}^b\Delta_{mn}^a}{(\omega_{ml}-\omega)^2}))
\end{eqnarray}
is the contribution of the mixed interband and intraband processes.
Here $r_{nm}^a$ and $r_{nm;b}^a$ are the $a$-component of the interband position matrix element and its generalized
momentum derivative, respectively. $\omega_{mn} = (\epsilon_{m{\bf k}} - \epsilon_{n{\bf k}})/\hbar$ where
$\epsilon_{m{\bf k}}$ is the $m$th band energy at the $\bf k$ point. $f_{mn} = f(\epsilon_{m{\bf k}}) - f(\epsilon_{n{\bf k}})$
where $f(\epsilon_{m{\bf k}})$ is the Fermi function. $\varepsilon_0$ and $V_c$ are the
vaccum permitivity and unit cell volume, respectively.

Linear electric-optic (LEO) effect refers to the linear refractive index variation ($\Delta n$)
with the applied electric field strength, $\Delta n = n^3r\mathcal{E}/2$, where $n$ is
the refraction index and $r$ is the LEO coefficient. The LEO effect thus
allows to exploit an electrical signal to control the amplitude, phase or direction
of a light beam in the NLO material, leading to applications in
high-speed optical modulation and sensing devices~\cite{Wu1996}.
In the zero frequency limit, the LEO coefficient is given by
\begin{equation}
r_{abc}(0) = -\frac{2}{\varepsilon _a(0)\varepsilon _b(0)} \lim_{\omega\to 0} \chi^{(2)}_{abc}(-2\omega,\omega,\omega).
\end{equation}
The LEO coefficient obtained from the SHG spectrum is referred as the electronic contribution.
There are also ionic and piezoelectric contributions to the LEO effect,\cite{veithen-prb2005}
which are out of scope of the present paper.

Another interesting second-order NLO response is 
the generation of DC photocurrents~\cite{Sipe2000,Nastos2010}.
These photocurrents are the main contributions to the bulk photovoltaic effect (BPVE). 
In the IPA, these photocurrents are given by~\cite{Sipe2000,Nastos2010} 
\begin{equation}
J_a(0) = \sigma^{abc}(0;\omega,-\omega)\mathcal{E}_b(\omega) \mathcal{E}_c(-\omega).
\end{equation}
In a nonmagnetic solid, the real part of the photoconductivity $\sigma^{abc}$ 
gives rise to the photocurrent due to purely linearly polarized light,
known as linear shift current, while the imaginary part produces 
photocurrent due to purely circularly polarized light, known as circular 
injection current~\cite{Sipe2000,Nastos2010}. However, in a $PT$-symmetric magnetic
material such as the AF CrI$_3$ BL, both the linear shift current and circular injection
currents would be zero~\cite{Ahn2020}. Instead, there are so-called circular shift 
current and linear injection current. 
The inject photocurrent conductivity $\sigma^{abc}$ for a magnetic system 
is given by ~\cite{Aversa1995,Ahn2020} $\sigma^{abc}=\tau \eta_{abc}$
where $\tau$ is the photoexcited carrier relaxation time and 
\begin{equation}
\eta_{abc}=\frac{-2\pi e^3}{\hbar^2 V_c}\sum _{\bf{k}} 
\sum_ {n,m}f_{nm}\Delta_{mn}^ar^b_{mn}r^{c}_{nm}\delta(\omega_{mn}-\omega).
\end{equation}
The shift photocurrent conductivity for a magnetic solid
is given by ~\cite{Aversa1995,Ahn2020} 
\begin{eqnarray}
\sigma_{abc}=&\frac{-i\pi e^3}{\hbar^2 V_c}\sum _{\bf{k}}
\sum_ {n,m}f_{nm}(r^b_{mn}r^{c}_{nm;a}-r^c_{nm}r^{b}_{mn;a}) \nonumber\\ 
&\delta(\omega_{mn}-\omega).
\end{eqnarray}

Since a large number of $k$-points are needed to get accurate NLO responses~\cite{Guo2005,Guo2008,Wu2008,Wang2015}, 
we use the efficient Wannier interpolation method based on maximally localized Wannier functions
(MLWFs)~\cite{WangX06,MarzariN,Ibanez-Azpiroz2018}. 
Total 112 MLWFs per unit cell of Cr $d$ and I $p$ orbitals are constructed by fitting to the
GGA+U+SOC band structure. The band structure obtained by the Wannier
interpolation is nearly identical to that from the GGA+U+SOC calculation. 
The SHG susceptibility and shift current conductivity are then evaluated by
taking a very dense $k$-point mesh of  200 $\times 200$ $\times$ 1.
The unit-cell volume $V_c$ in Eqs. (2), (3) and (6) is not well-defined for a two-dimensional (2D) system. 
Therefore, following the previous studies,\cite{Fang2018,vkg-njp2019}, we used the effective 
unit-cell volume of the BL rather than the volume of the supercell which is arbitrary.

\section{Results and discussion}
We first study the magnetic properties of the AF CrI$_3$ BL. 
The calculated spin magnetic moment on each Cr atom is $\pm$3.21 $\mu_B$, 
being in good agreement with the experimental value of $\sim$3.0 $\mu_B$  ~\cite{mag2015}.
The calculated magnetic moment of Cr is also consistent with three unpaired electrons in its $t_{2g}$ 
configuration in this structure. 
We also find significant proximity-induced magnetic moment of $\pm$0.09 $\mu_B$ on the I ion. 
Nevertheless, the sum of all the magnetic moments on the I and Cr sites in the unit cell is zero.

\begin{figure}[htb]
\begin{center}
\includegraphics[width=7cm]{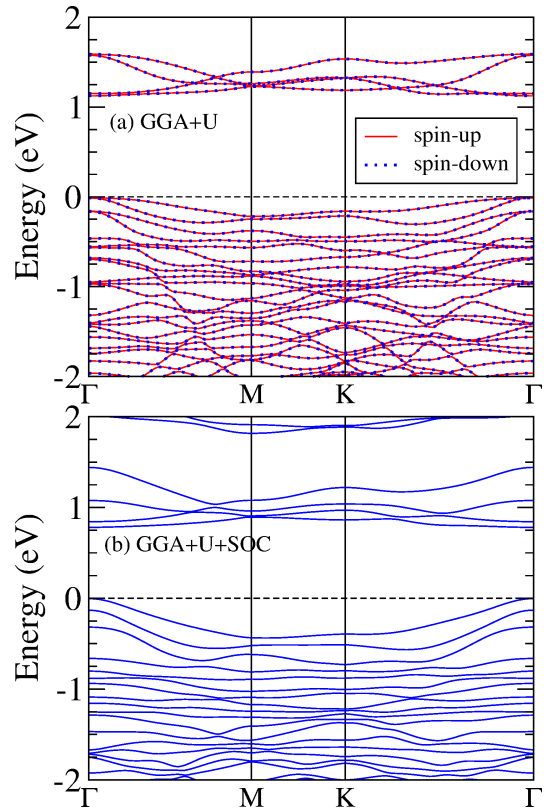}
\end{center}
\caption{(a) Spin-polarized scalar-relativistic and (b) fully relativistic  band structures of BL CrI$_3$. 
Horizontal dashed-lines denote the top of valance band.}
\end{figure}

\subsection{Electronic structure}
A better understanding of the optical properties can be achieved through a detailed analysis 
of the band structure. Therefore, we display scalar-relativistic (Fig. 2(a)) and  fully relativistic (Fig. 2(b))
band structures of BL CrI$_3$ in Fig. 2. 
Overall, the scalar-relativistic and fully relativistic band structures look rather similar.
In particular, all the bands remain doubly degenerate, i.e., Kramers degeneracy is maintained.
In the scalar-relativistic case, this is consistent with the antiferromagnetism in a centrosymmetric crystal.
In the fully-relativistic case, this is somewhat surprising because the AF configuration
here breaks both $T$ symmetry and $P$ symmetry~\cite{sun-nature2019}.
Nevertheless, further analysis indicates that AF BL CrI$_3$ has the $PT$ symmetry \cite{Zhang2019a}, 
which protects the Kramers degeneracy.
However, the energy positions and shapes of some energy bands do change significantly when the SOC is included. 
Although Fig. 2 shows that both band structures are a semiconductor with the direct band gap 
located at the $\Gamma$ point, the fully relativistic band struture has a band gap of 0.78 eV,
being smaller than that of the scalar-relativistic one (1.04 eV). 
The most important effect of the SOC is that the NLO responses of AF BL CrI$_3$ studied in this paper
occur only when the SOC is taken into account. This is because the electrons in AF BL CrI$_3$ would
only feel the broken  $P$-symmetry caused by the AF configuration when the SOC is included. 
Therefore, all the optical properties of AF BL CrI$_3$ presented below are calculated from
the fully relativistic band structure. 

We present the total as well as site- and orbital-projected densities of states (DOS) for BL CrI$_3$ in Fig. 3. 
It is clear from Fig. 3 that the valance and conduction bands in the energy ranges between -3.0 and 0.0 eV as well as
between 1.0 eV and 3.0 eV, respectively, are dominated by the Cr $d$ orbitals with a significant contribution 
from the I $p$ orbitals, suggesting the strong hybridization between Cr $d$ and I $p$ orbitals in the CrI$_3$ structure. 
Specifically, the upper valance band region between -1.75 eV and  0.0 eV are made up of Cr $t_{2g}$ 
(i.e., $d_{xy,x^2-y^2}$ and $d_{z^2}$) orbitals. The lower conduction bands ranging from 1.0 eV to 1.6 eV 
stem mainly from the Cr $e_g$ (i.e., $d_{xz,yz}$) orbitals (see Fig, 3(b)). This indicates that the band gap in the CrI$_3$ 
structure is caused by the crystal-field splitting of the Cr $t_{2g}$ and $e_{g}$ bands. 
The conduction bands ranging from 1.75 eV to 3.0 eV are made up  mainly of the 
Cr $d_{xy,x^2-y^2}$ and $d_{z^2}$ orbitals.  

It is well-known that the band gap calculated based on the GGA functional is usually too small
compared with the experimental one, mainly because the many-body effects especially quasiparticle 
self-energy correction are not fully taken into account in the GGA. For example,
the GGA band gap of bulk  CrI$_3$ is 0.62 eV~\cite{vkg-njp2019} 
being about 50 \% smaller than the experimental value of 1.2 eV~\cite{dillon1965}.
Nonetheless, semilocal hybrid Heyd-Scuseria-Ernzerhof (HSE) functional~\cite{heyd2003j,heyd2006} 
is known to produce improved band gaps for semiconductors. 
We have performed the HSE calculations for bulk and multilayer CrI$_3$ using the HSE06 functional~\cite{vkg-njp2019}.
Indeed, we obtained the HSE band gap of 1.22 eV for bulk CrI$_3$, 
being in excellent agreement with the experimental value.
Therefore, in the present calculations of the optical and NLO properties, 
we use the HSE band gap of 1.33 eV for BL CrI$_3$ and then adopt the  scissors correction (SC) scheme.

\begin{figure}[htb]
\begin{center}
\includegraphics[width=7cm]{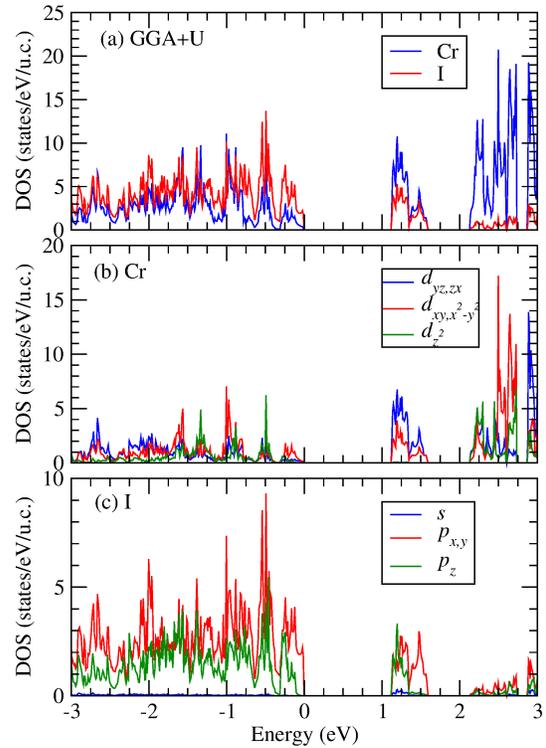}
\caption{Site- and orbital-projected densities of states (DOS) of BL CrI$_3$ from
the scalar-relativistic calculation.}
\end{center}
\end{figure}

\begin{figure}[htb]
\begin{center}
\includegraphics[width=7cm]{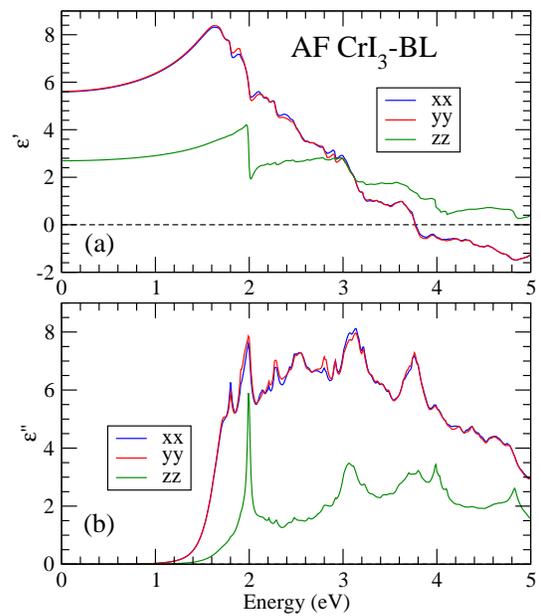}
\end{center}
\caption{(a) Real and (b) imaginary parts of the dielectric function of BL CrI$_3$.}	
\end{figure}

\subsection{Dielectric function}
To help understand the calculated SHG and photocurrent of BL CrI$_3$ 
and also to estimate the linear electro-optical effect, we also calculate
the linear optical properties. 
The calculated real ($\varepsilon'(\omega)$) (dispersive) and imaginary ($\varepsilon''(\omega)$) 
(absorptive) part of the optical dielectric function of the BL CrI$_3$ are presented in Fig. 4. 
Notably, the dielectric function of out-of-plane electric polarization $\varepsilon_{zz}$ ($\mathcal{E}||c$)
differs significantly from that of two in-plane electric polarizations $\varepsilon_{xx}$ and $\varepsilon_{yy}$
(note that $x||a$), although the latter two are nearly identical (see Fig. 4).
This strong optical anisotropy is expected from the 2D nature of the BL CrI$_3$ structure.
Specifically, in the case of the real part of the dielectric constant,
the two in-plane polarization ones ($\varepsilon'_{xx}$ and $\varepsilon'_{yy}$)
are nearly twice as large as that of the out-of-plane electric polarization one ($\varepsilon'_{zz}$)
in the photon energy from 0.0 to 2.0 eV. As the photon energy further increases, this difference
steadily reduces and at $\sim$3.0 eV or above, $\varepsilon'_{zz}$ becomes larger
than $\varepsilon'_{xx}$ and $\varepsilon'_{yy}$ [see Fig. 4(a)].
In the case of the imaginary part of the dielectric constant, $\varepsilon''_{xx}$ and $\varepsilon''_{yy}$
start to increase sharply at $\sim$1.4 eV until 2.0 eV, while $\varepsilon''_{zz}$ starts to
increase rapidly only at $\sim$1.8 eV. In the photon energy range from 2.0 eV to 5.0 eV,
$\varepsilon''_{xx}$ and $\varepsilon''_{yy}$ are about two-times larger than $\varepsilon''_{zz}$.

This strong anisotropy in the linear optical property of bulk and multilayers CrI$_3$  
was noticed before in, e.g., the previous study of Ref. \cite{vkg-njp2019}. 
This behavior of the linear optical property can be understood in terms of 
the calcated orbital-decomposed DOS presented in the proceeding section. 
In the energy range of -3 to -0.3 eV in the valence band region, 
the enery bands are dominated by the Cr $d$ orbitals [see Fig. 3(b)], with a major contribution 
from the $d_{xy,x^2-y^2}$ orbitals. It was shown \cite{vkg-njp2019}  that $d_{xy,x^2-y^2}$ ($d_{z^2}$) states 
can be excited by only $\mathcal{E}\perp c$ ($\mathcal{E}\parallel c$) polarized light while Cr $d_{xz,yz}$ 
orbitals can be excited by light of both polarization. This suggests that $\varepsilon'_{xx}$ usually 
is greater than $\varepsilon'_{zz}$ in the low energy region (see Fig. 4).

\subsection{Second-harmonic generation}
Symmetry considerations allow us to identify nonzero elements of the third-rank SHG 
susceptibility tensor. The AB$'$ stacked BL CrI$_3$ has 
the centrosymmetric $C_{2h}$ ($2/m$) space group and consequently
should not have any second-order NLO responses. 
As mentioned before, however, the AF structure breaks the spatial inversion and time-reversal symmetries,
resulting in a magnetic space group of $C_2$ ($2'$). The $C_2$ magnetic space group has ten nonvanishing elements
in the SHG susceptibility tensor \cite{Gallego2019}. By considering the elements of 
the SHG susceptibility with only the $xy$-plane polarizations (i.e., considering only the normal incidence), 
we are left with only three independent non-zero elements, namely, $\chi^{(2)}_{xxx}$, 
$\chi^{(2)}_{xyy}$, $\chi^{(2)}_{yxy}$.
The calculated real, imaginary and absolutes values of these three non-vanishing elements of BL CrI$_3$
are presented in Fig. 5. 

As expected, Fig. 5(b) shows that the imaginary (absorptive) part of the $\chi^{(2)}$ spectra of
all the three elements is zero for photon energy ($\hbar\omega$) below 0.67 eV (mid-band gap).
As $\hbar\omega$ further increases, the imaginary part increases sharply in magnitude and reaches
its maximum just above 0.8 eV. After that, the magnitudes of the imaginary part of all the elements
decreases oscillatorily all the way up to 5.0 eV.
The magnitude of the real part and absolute value of the $\chi^{(2)}$ spectra
are not zero at $\hbar\omega=0$ and increases gradually with $\hbar\omega$ up to $\sim$0.67 eV. 
When $\hbar\omega$ further increases, they increases dramatically and reaches the maximum at $\sim$0.8 eV (see Fig. 5(a)).
Among the three elements, we find that the magnitude of the imaginary part and absolute value
of $\chi^{(2)}_{xxx}$ is more or less larger than the other two elements in the energy
range from 0.0 to 3.0 eV (see Fig. 5(c)). 
In particular, the maximum absolute value of the $\chi^{(2)}_{xxx}$ is as large as $\sim$350 pm/V
at $\hbar\omega=0.85$ eV.
We also notice that $\chi^{(2)}_{xyy}$ has a sharp narrow peak with its absolute value of
$\sim$350 pm/V at 1.7 eV.

\begin{figure}[htb]
\begin{center}
\includegraphics[width=7cm]{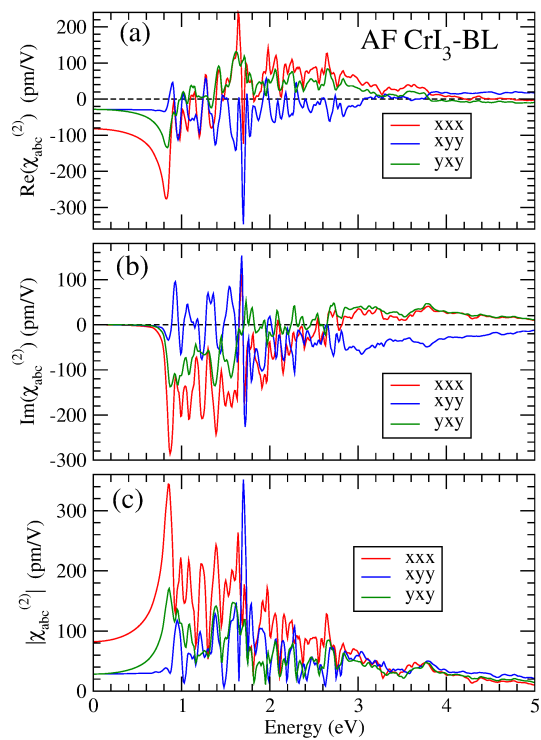}
\end{center}
\caption{(a) Real, (b) imaginary and (c) absolute value of the SHG spectra of BL CrI$_3$.}	
\end{figure}

\begin{table*}[htbp]
\begin{center}
\caption{Calculated dielectric constant ($\varepsilon_{yy}\approx \varepsilon_{xx}$), second-order optical susceptibility ($\chi_{abc}^{(2)}$), 
and linear electro-optical coefficient ($r_{abc}$) of BL CrI$_3$. }
\begin{tabular}{cccccccccc}
\hline \hline
 $\hbar \omega$ &$\varepsilon_{xx}$ &$\varepsilon_{zz}$ &$\chi^{(2)}_{xxx}$ &$\chi^{(2)}_{xyx}$ &$\chi^{(2)}_{yxy}$ &$r_{xxx}$ &$r_{xyx}$    &$r_{yxy}$  \\
     (eV)       &               &                &       (pm/V)      &    (pm/V)         &      (pm/V)       &   (pm/V) &   ((pm)/V)  &  (pm/V) \\
\hline
0.0  	&   5.60  &  2.69       & -82.5	         &  -28.5   & -28.5  &    5.26		& 1.82	& 1.82 \\
0.7     &   5.93  &  2.79       &  159           &   31.9   &  67.6  &   10.1		& 2.03  & 4.31 \\
\hline \hline
\end{tabular} \\
\end{center}
\end{table*}

To analyze the origins of the prominent features in the calculated 
SHG spectra, we display the modulus of the imaginary part and absolute value 
of the second-order susceptibilities $|\chi^{(2)} (-2\omega,\omega,\omega)|$ 
together with the imaginary part of the dielectric functions $\varepsilon''(\omega)$ 
and $\varepsilon''(2\omega)$ in Fig. 6. 
Figure 6(a) and (b) shows that the peaks in the $|Im[\chi^{(2)} (-2\omega,\omega,\omega)]|$ 
in the energy range from the mid-band gap ($\sim$0.67 eV) to the absorption edge ($\sim$1.33 eV) 
can be correlated with the features in the $\varepsilon''(2\omega)$ spectra (see Fig. 6(c)), 
indicating that they are due to double-photon (2$\omega$) resonances. 
The peaks above the absorption edge, on the other hand, 
can be related to the features in either the $\varepsilon''(2\omega)$ or $\varepsilon''(\omega)$ or even both, 
suggesting that they can be caused by both double-photon and single-photon resonance. 
Because of the contributions from both one-and two-photon resonances, the spectra oscillate rapidly 
in this region and diminish gradually at higher photon energies. 
Both the spectrum of $\chi^{(2)}_{xxx}$ and $\chi^{(2)}_{xyy}$ 
elements have almost the same oscillatory behavior. 
\vspace{10pt}

\begin{figure}[htb]
\begin{center}
\includegraphics[width=7cm]{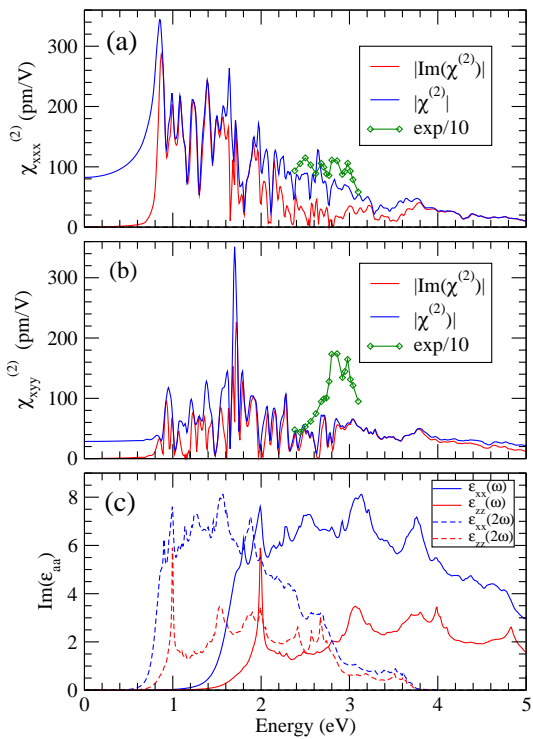}
\end{center}
\caption{Modulus and absolute value of the imaginary part of the second-order susceptibility 
of (a) $\chi^{(2)}_{xxx}$ and (b) $\chi^{(2)}_{xyy}$ of BL CrI$_3$. (c) Imaginary part of the dielectric 
function $\varepsilon"$ of BL CrI$_3$. Solid line with green diamonds  
in (a) and (b) represents the experimental SHG spectrum from Ref. ~\cite{sun-nature2019}. 
The experimental SHG spectra are scaled by a factor of 1/10.}	
\end{figure}

The SHG coefficients of some bulk AF materials such as the classic $PT$-symmetric Cr$_2$O$_3$
have been measured before. We notice that compared with these bulk materials, the magnetic SHG coefficients
of the BL CrI$_3$ are giantic, being two-orders of magnitude larger than that
of Cr$_2$O$_3$ ($\sim$2 pm/V) \cite{sun-nature2019}.
Furthermore, we notice that the surface/interface-induced SHG in
the ferromagnetic thin-films is even smaller, being in the order of 0.4$\times$ 10$^{-7}$ pm/V \cite{pan-prb1989,reif-prl1991}.
The magnetic SHG found in the BL CrI$_3$ is nonreciprocal and also the sign of the SHG is controllable
by switching the magnetization direction~\cite{sun-nature2019}. Therefore,
the large SHG found in the BL CrI$_3$ suggests
that the BL CrI$_3$  will have valuable applications
in magneto-optical NLO devices.

Let us now compare the SHG of the present material with that of other 2D materials.
First we notice that the modulus of $\chi^{(2)}_{xxx}$ at zero frequency 
in BL CrI$_3$ is two times larger than $\chi^{(2)}_{yyy}$ of the BN sheet~\cite{Guo2005,Guo2008},
although it is nearly three times smaller than $\chi^{(2)}_{yyy}$ of the SiC sheet~\cite{Wu2008}.
Wang and Guo~\cite{Wang2015} recently performed systematic {\it ab initio} studies of the second-order
NLO responses of odd number few-layers of group 6B transition metal dichalcogenides.  
The spatial inversion symmetry in these odd number few-layer structures is broken and
thus gives rise to large SHG coefficient and also significant LEO effect~\cite{Wang2015}. 
Strikingly, the magnetic SHG spectra of the AF CrI$_3$ BL (Figs. 5 and 6)
are in the same order of magnitude as that of monolayer MoS$_2$~\cite{Wang2015}, 
which is the best candidate of the NLO material among the 2D materials \cite{li-prb2013,kumar-prb2013,malard-prb2013}. 
This demonstrates that the magnetism-induced SHG in a centrosymmetric crystal 
can be as large as that of a crystal with the broken $P$ symmetry. 
Thus the excellent nonreciprocal NLO properties of the AF BL CrI$_3$ will find promising
applications in ultrathin light signal modulators, second-harmonic and sum-frequency generation
devices and in electro-optical switches.

As mentioned above, the SHG experiments on the AF CrI$_3$ BL have been very recently carried out \cite{sun-nature2019}.
The experimental $\chi^{(2)}_{xxx}$ and $\chi^{(2)}_{xyy}$ spectra are re-plotted, respectively, in Figs. 6(a) and 6(b) for comparison. 
Figure 6 shows that overall, the peak positions and line shapes of the calculated SHG spectra are in good agreement 
with the experimental spectra.
Nonetheless, the measured SHG spectra are generally one-order of magnitude larger than the calculated spectra.
The discrepancy in the magnitude of the spectra between the experiment and the present theory 
might be due to the fact that in the present calculation a free-standing  CrI$_3$ BL is considered, 
whereas in the experiment \cite{sun-nature2019} the mechanical exfoliated CrI$_3$ was put on the substrate of Si/SiO$_2$ 
and was also encapsulated by hBN thin flakes \cite{sun-nature2019}. 
We notice that the sample preparation of a 2D material can have a significant 
effect on the optical and NLO responses of the 2D material. 
For example, the presence of the interface between the substrate and the material may
break the inversion symmetry of the substrate and thus induces additional non-vanishing SHG 
spectra \cite{guyot-prb1986,trassin-am2015,matsubara-sc2015}. 
This suggests that further SHG measurements on the AF BL CrI$_3$ structure are needed to resolve this discrepancy.
Moreover, the experimental reported $\chi^{(2)}$ values of ML MoS$_2$ 
can vary as much as three-orders of magnitude. The experiental $|\chi^{(2)}|$ at 810 nm wavelength of the mechanically
exfoliated ML MoS$_2$ reported in Ref. ~\cite{guyot-prb1986} is as large as $\sim 10^5$ pm/V,
while the ML-MoS$_2$ prepared by chemical vapor deposition has the $\chi^{(2)}$
value of 5$\times$10$^3$ pm/V \cite{kumar-prb2013} and in another experiment the value of $\chi^{(2)}$
was reported to be about 320 pm/V \cite{li-prb2013}. 

\subsection{Linear electro-optical coefficient}
The LEO effect has been used extensively in the integrated optical devices 
for optical communications and modulating light. Thus, we also calculate the LEO coefficient 
of BL CrI$_3$ using the obtained dielectric constants and SHG susceptibility [see Eq. (4)].
The calculated LEO coefficients $r_{abc}(0)$ at photon energy of zero and 0.7 eV, along with 
the corresponding dielectric constants and SHG coefficients are presented in Table 1. 
Interestingly, there is a strong in-plane anistropy in both the SHG and LEO coefficients (see Table 1). 
Furthermore, all the static $r_{abc}$ values of BL CrI$_3$ are significantly larger than
the corresponding elements of $r_{abc}$ of the MLs of transition metal dichalcogenide semiconductors~\cite{Wang2015}.
In particular, the modulus of $r_{xxx}$ of BL CrI$_3$ (5.26 pm/V) is 
more than four times larger than that of $r_{xxy}$ of ML MoS$_2$ (1.23 pm/V)~\cite{Wang2015}.
We also notice that this value is three times larger than that of the BN sheet~\cite{Guo2005,Guo2008}.
This is somewhat surprising since the magnetism-induced second-order NLO responses are
expected to be smaller than that due to the broken inversion symmetry in the crystal structure.
Thus, the significant nature of the magnetism-induced SHG spectra 
and LEO coefficients of BL CrI$_3$ would enable it to find promising applications 
in magnetic material-based optical communications.

\begin{figure}[htb]
\begin{center}
\includegraphics[width=7cm]{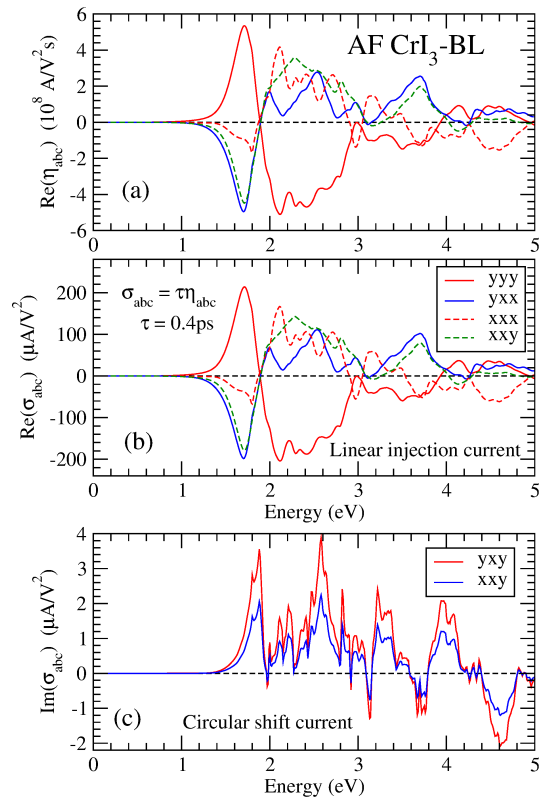}
\end{center}
\caption{(a) Injection photocurrent susceptibility $\eta_{abc}$ and (b) conductivity $\sigma_{abc}$ 
of BL CrI$_3$. (c) Shift photocurrent conductivity of BL CrI$_3$.}
\end{figure}

\subsection{Shift and injection photocurrents}
As discussed above in Sec. 2, in the AF BL CrI$_3$, because of its $PT$ symmetry,
there is no nonvanishing element of usual linear shift photocurrent tensor and circular
injection photocurrent which occur in nonmagnetic broken $P$-symmetry materials~\cite{Ahn2020}.
However, there are nonzero elements of circular shift current and linear injection current instead.
Here we consider only the in-plane polarizations of optical electric field and the in-plane 
photocurrents. Consequently, we have six independent nonzero elements 
of  the linear injection current tensor~\cite{Gallego2019}, namely, $xxx, xyy, xxy, yxx, yyy$ and $yxy$.
We display four pronounced nonzero elements of the calculated injection 
current susceptibility $\eta_{abc}$ in Figs. 7(a).
Note that in the AB-stacked AF BL CrI$_3$, there are only two independent in-plane nonzero elements 
of the linear injection current tensor,~\cite{Zhang2019a} namely,  $\sigma_{xxx}$ and $\sigma_{yyy}$.
Using the typical relaxation time of $\tau = 0.4$ps~\cite{Zhang2019a}, we obtain the injection photocurrent conductivity
$\sigma_{abc}$ [see Eq.(6) in Sec. 2], as shown in Fig. 7(b). 
There are also two independent in-plane nonzero elements of circular shift current tensor,
as displayed in Fig. 7(c). In contrast, there is no nonzero element of circular shift current tensor
in the AB-stacked AF BL CrI$_3$~\cite{Zhang2019a}. These additional nonzero elements
of the photoconductivity tensors in the AB$'$-stacked CrI$_3$ BL studied here will allow one to
characterize the structure of the CrI$_3$ BL by the BPVE experiments.

Overall, the spectra of all the four $\sigma_{abc}$ ($\eta_{abc}$) elements displayed 
in Fig. 7(b) [7(a)] have quite similar shapes and magnitudes.
However, the $yyy$ spectrum has the opposite sign to that of the rest three spectra 
over almost the entire energy range.
They are all zero below the band gap of 1.33 eV and they increase rapidly above the absorption edge.
In particular, $\sigma_{yyy}$, $\sigma_{yxx}$ and $\sigma_{xxy}$ reach their maximum of $\sim$210 $\mu$A/V$^2$ at $\sim$1.71 eV.
As the photon energy further increases, $\sigma_{yyy}$ decreases rapidly, changing sign at $\sim$1.89 eV
and reaches its negative maximum of -204 $\mu$A/V$^2$ at 2.12 eV.
We notice that our $yxx$ linear injection current spectrum is similar in both magnitude
and shape to the $yyy$ spectrum of the AB-stacked CrI$_3$ BL reported
by Zhang {\it et al.}~\cite{Zhang2019a}. The magnitude of our $yyy$ spectrum is
similar to that of both $xxx$ and $yyy$ spectra in~\cite{Zhang2019a}.

The two circular shift current spectra are almost identical in shape [see Fig. 7(c)].
The only difference is that the $xxy$ spectrum appears to be a down-scaled $yxy$ spectrum.
For example, they both increase rapidly above the absorption edge of 1.33 eV
and then peak at $\sim$1.9 eV. As photon energy further increases, the two spectra
decrease sharply and become nearly zero at $\sim$2.0 eV. After that, they increase again
with photon energy. The circular shift currents [Fig. 7(c)] appear to be two-orders of magnitude 
smaller than the linear injection currents [Fig. 7(b)]. Nonetheless, the size of
the circular shift currents is comparable to that of the linear shift currents in 
archetypal semiconductors with a similar band gap such as CdSe~\cite{Nastos2010}.

Let us now compare the photocurrents of BL CrI$_3$ with the well-known BPVE materials 
to examine the possibility of BL CrI$_3$ for photovoltaic solar cell applications.
The predicted shift photocurrent conductivity of the well-known ferroelectric materials of PbTiO$_3$ and BiTiO$_3$ 
have a maximum value of 5 $\mu$A/V$^2$ within the visible region below 6 eV.
This is much smaller than that of BL CrI$_3$.
Nevertheless, the predicted shift current conductivity of the chiral materials 
has the highest value of $\sim$80 $\mu$A/V$^2$ in the visible light \cite{Zhang2019b}. 
The experimental reported maximum value of the shift photocurrent conductivity within the visible range 
is 6 $\mu$A/V$^2$ for semiconductor SbSI \cite{Sotome2019}. 
All these comparisons indicate that the magnetism-induced photocurrent in BL CrI$_3$ is very large.
Furthermore, compared with ferroelectric oxides such as PbTiO$_3$ and BiTiO$_3$,
BL CrI$_3$ has a much smaller band gap and thus has a high solar energy absorption efficiency.
Therefore, BL CrI$_3$ will be a promising material for high efficient photovoltaic solar cell application.

\section{Conclusion}
In conclusion, we have calculated AF-induced NLO responses of
the AB$'$-stacked centrosymmetric BL CrI$_3$ based on the GGA+U method.
Strikingly, we find that the magnetic SHG, LEO and photocurrent in the AF 
BL CrI$_3$ are very large, being comparable or even larger than
the well-known nonmagnetic noncentrosymmetric semiconductors with the same properties.
For example, the calculated SHG coefficients are in the same order of magnitude
as that of MoS$_2$ ML, the most promising 2D material for NLO devices.
The calculated LEO coefficients are almost three times larger than
that of MoS$_2$ ML. The calculated NLO photocurrent in BL CrI$_3$
is among the largest values predicted so far for the bulk photovoltaic materials.     
On the other hand, unlike nonmagnetic semiconductors, the NLO responses in the AF BL CrI$_3$ are
nonreciprocal and also switchable by rotating magnetization direction.
Therefore, our interesting findings show that the AF BL CrI$_3$ will 
find valuable applications in magnetic 
NLO and LEO devices such as frequency conversion, electro-optical switches, 
and light signal modulators as well as high energy conversion efficiency
photovoltaic solar cells.

{\it Note added --} After submitting our manuscript, we became aware of
the work by Song {\it et al.},~\cite{Song2020} who independently performed 
the {\it ab initio} calculation on the SHG in the AF BL CrI$_3$ based 
on the velocity gauge formalism instead of length gauge formalism adopted here.

\textbf{Acknowledgement:}
G.-Y. Guo thanks Junyeong Ahn and Naoto Nagaosa for helpful discussions on
NLO photocurrent and bulk photovoltaic effect.
The authors acknowledges the support by the Ministry of Science and Technology, National Center for 
Theoretical Sciences and National Center for High-performance Computing in Taiwan. 
G.-Y. Guo also thanks the support from the Far Eastern Y. Z. Hsu Science and Technology
Memorial Foundation in Taiwan.

\textbf{References:}
\label{refs}
{}
\end{document}